\documentclass{article}

\usepackage{microtype}
\usepackage{graphicx}
\usepackage{caption}
\usepackage{subcaption}
\usepackage{booktabs} %
\usepackage{bm} %

\usepackage{hyperref}

\usepackage[accepted]{icml2023}

\usepackage{amsmath}
\usepackage{amssymb}
\usepackage{mathtools}
\usepackage{amsthm}
\usepackage{enumitem}
\usepackage[capitalize,noabbrev]{cleveref}

\theoremstyle{plain}

\theoremstyle{definition}

\theoremstyle{remark}

\usepackage{etoolbox}

\icmltitlerunning{MusicLM: Generating Music From Text}

\begin{document}

\twocolumn[
\icmltitle{MusicLM: Generating Music From Text}

\icmlsetsymbol{equal}{*}

\begin{icmlauthorlist}
\icmlauthor{Andrea Agostinelli}{equal,google}
\icmlauthor{Timo I. Denk}{equal,google}\\
\icmlauthor{Zalán Borsos}{google}
\icmlauthor{Jesse Engel}{google}
\icmlauthor{Mauro Verzetti}{google}
\icmlauthor{Antoine Caillon}{ircam}
\icmlauthor{Qingqing Huang}{google}
\icmlauthor{Aren Jansen}{google}
\icmlauthor{Adam Roberts}{google}
\icmlauthor{Marco Tagliasacchi}{google}
\icmlauthor{Matt Sharifi}{google}
\icmlauthor{Neil Zeghidour}{google}
\icmlauthor{Christian Frank}{google}
\end{icmlauthorlist}

\icmlaffiliation{google}{Google Research}
\icmlaffiliation{ircam}{IRCAM - Sorbonne Université (work done while interning at Google)}

\icmlcorrespondingauthor{Christian Frank}{chfrank@google.com}

\icmlkeywords{music, generative modeling, machine learning, text-conditioning, autoregressive}

\vskip 0.3in
]

\printAffiliationsAndNotice{\icmlEqualContribution} %

\providetoggle{showcomments}
\settoggle{showcomments}{true} 								%

\iftoggle{showcomments}{%
    \newcommand{\changed}[1]{\textcolor{blue}{#1}}
    \newcommand{\todo}[1]{\textcolor{blue}{\textbf{TODO:} #1}}
    \newcommand{\resolved}[3][]{\ifstrequal{#1}{resolved}{\textcolor{blue}{RESOLVED:}~\textbf{{\MakeUppercase #2:}}~{#3}}{\textbf{\MakeUppercase #2:}~#3}}
    \newcommand{\andrea}[2][]{\textcolor{brown}{\resolved[#1]{andrea}{#2}}}
    \newcommand{\timo}[2][]{\textcolor{ForestGreen}{\resolved[#1]{timo}{#2}}}
    \newcommand{\christian}[2][]{\textcolor{violet}{\resolved[#1]{christian}{#2}}}
    \newcommand{\jesse}[2][]{\textcolor{DarkYellow}{\resolved[#1]{jesse}{#2}}}
    \newcommand{\assign}[2][]{\textcolor{red}{\resolved[#1]{Assigned to: }{#2}}}
}{%
    \newcommand{\changed}[1]{#1}
    \newcommand{\todo}[1]{}
    \newcommand{\andrea}[2][]{}
    \newcommand{\timo}[2][]{}
    \newcommand{\christian}[2][]{}
}

\definecolor{ForestGreen}{RGB}{34,139,34}
\definecolor{DarkYellow}{RGB}{204,204,0}

\newcommand{\model}{MusicLM} %
\newcommand{\modelmv}{MusicLM-mv}
\newcommand{\dataset}{MusicCaps}
\newcommand{\datasetshort}{MusicCaps}
\newcommand{\mulan}{MuLan}
\newcommand{\audiolm}{AudioLM}
\newcommand{\soundstream}{SoundStream}
\newcommand{\dalletwo}{DALL$\cdot$E~2}
\newcommand{\mulanText}{M_T}
\newcommand{\mulanAudio}{M_A}
\newcommand{\sounds}{A}
\newcommand{\mlm}{S}

\newcommand{\mlmcodebook}{K}
\newcommand{\matchingthreshold}{\tau}
\newcommand{\memorizationpromptlen}{T}

\newcommand{\website}{https://goto.google.com/text2music-website}

\linepenalty=1000  %

\begin{abstract}

We introduce {\model}, a model for generating high-fidelity music from text descriptions such as \textit{``a~calming violin melody backed by a distorted guitar~riff''}. {\model} casts the process of conditional music generation as a hierarchical sequence-to-sequence modeling task,
and it generates music at 24~kHz that remains consistent over several minutes.
Our experiments show that {\model} outperforms previous systems both in audio quality and adherence to the text descriptions. Moreover, we demonstrate that {\model} can be conditioned on both text and a melody in that it can transform whistled and hummed melodies according to the style described in a text caption. To support future research, we publicly release {\dataset}, a dataset composed of 5.5k~music-text pairs, with rich text descriptions provided by human experts.

\small{\href{https://google-research.github.io/seanet/musiclm/examples/}{google-research.github.io/seanet/musiclm/examples}}

\end{abstract}

\section{Introduction}
\label{intro}

\looseness=-1
Conditional neural audio generation covers a wide range of applications, ranging from text-to-speech~\citep{ze2013statistical, oord2016wavenet} to lyrics-conditioned music generation~\cite{jukebox} and audio synthesis from MIDI sequences~\citep{hawthorne22_diffusion}. Such tasks are facilitated by a certain level of temporal alignment between the conditioning signal and the corresponding audio output. In contrast, and inspired by progress in text-to-image generation~\cite{dalle, dalle-2, imagen, parti}, recent work has explored generating audio from sequence-wide, high-level captions~\citep{yang2022diffsound, audiogen} such as \textit{``whistling with wind blowing''}. While generating audio from such coarse captions represents a breakthrough, these models remain limited to simple acoustic scenes, consisting of few acoustic events over a period of seconds. Hence, turning a single text caption into a rich audio sequence with long-term structure and many stems, such as a music clip, remains an open challenge.

\looseness=-1
\audiolm~\cite{audiolm} has recently been proposed as a framework for audio generation. Casting audio synthesis as a language modeling task in a discrete representation space, and leveraging a hierarchy of coarse-to-fine audio discrete units (or \textit{tokens}), \audiolm~ achieves both high-fidelity and long-term coherence over dozens of seconds. Moreover, by making no assumptions about the content of the audio signal, \audiolm~ learns to generate realistic audio from audio-only corpora, be it speech or piano music, without any  annotation. The ability to model diverse signals suggests that such a system could generate richer outputs if trained on the appropriate data.

\looseness=-1
 Besides the inherent difficulty of synthesizing high-quality and coherent audio, another impeding factor is the scarcity of paired audio-text data. This is in stark contrast with the image domain, where the availability of massive datasets contributed significantly to the remarkable image generation quality that has recently been achieved~\citep{dalle, dalle-2, imagen, parti}. Moreover, creating text descriptions of general audio is considerably harder than describing images. First, it is not straightforward to unambiguously capture with just a few words the salient characteristics of either acoustic scenes (e.g., the sounds heard in a train station or in a forest) or music (e.g., the melody, the rhythm, the timbre of vocals and the many instruments used in accompaniment). Second, audio is structured along a temporal dimension which makes sequence-wide captions a much weaker level of annotation than an image caption.

\looseness=-1
In this work, we introduce {\model}, a model for generating high-fidelity music from text descriptions. {\model} leverages {\audiolm}'s multi-stage autoregressive modeling as the generative component, while extending it to incorporate text conditioning. To address the main challenge of paired data scarcity, we rely on {\mulan} \citep{mulan}, a joint music-text model that is trained to project music and its corresponding text description to representations close to each other in an embedding space. This shared embedding space eliminates the need for captions at training time altogether, and allows training on massive audio-only corpora. That is, we use the {\mulan} embeddings computed from the audio as conditioning during training, while we use {\mulan} embeddings computed from the text input during inference.

When trained on a large dataset of unlabeled music, {\model} learns to generate long and coherent music at 24 kHz, for text descriptions of significant complexity, such as \textit{``enchanting jazz song with a memorable saxophone solo and a solo singer''} or \textit{``Berlin 90s techno with a low bass and strong kick''}. To address the lack of evaluation data for this task, we introduce {\dataset}, a new high-quality music caption dataset with 5.5k~examples prepared by expert musicians, which we publicly release to support future research.

Our experiments show through quantitative metrics and human evaluations that {\model} outperforms previous systems such as Mubert~\citep{Mubert} and Riffusion~\citep{riffusion}, both in terms of quality and adherence to the caption. 
Furthermore, since describing some aspects of music with words can be difficult or even impossible, we show how our method supports conditioning signals beyond text. Concretely, we extend {\model} to accept an additional melody in the form of audio (e.g., whistling, humming) as conditioning to generate a music clip that follows the desired melody, rendered in the style described by the text prompt.

\looseness=-1
We acknowledge the risks associated with music generation, in particular, the potential misappropriation of creative content. In accordance with responsible model development practices, we conduct a thorough study of memorization by adapting and extending the methodology of~\citet{carlini-memorization} used for text-based large language models. Our findings show that when feeding {\mulan} embeddings to {\model}, the sequences of generated tokens significantly differ from the corresponding sequences in the training set.

The key contributions of this work are the following:
\begin{enumerate}%
    \item We introduce {\model}, a generative model that produces high-quality music at 24~kHz which is consistent over several minutes while being faithful to a text conditioning signal.
    \item We extend our method to other conditioning signals, such as a melody that is then synthesized according to the text prompt. Furthermore, we demonstrate long and coherent music generation of up to 5-minute long clips.
    \item We release the first evaluation dataset collected specifically for the task of text-to-music generation: {\dataset} is a hand-curated, high-quality dataset of 5.5k~music-text pairs prepared by musicians.
\end{enumerate}

\label{contributions}
\newpage
\section{Background and Related Work}
\label{sec:background-related}

The state-of-the-art in generative modeling for various domains is largely dominated either by Transformer-based autoregressive models~\citep{vaswani2017attention} or U-Net-based diffusion models~\citep{ho2020denoising}. In this section, we review the related work with an emphasis on autoregressive generative models operating on discrete tokens, which share similarities with {\model}.

\subsection{Quantization} 

\looseness=-1
Modeling sequences of discrete tokens autoregressively has proven to be a powerful approach in natural language processing \citep{gpt3, lamda} and image or video generation \citep{esser2021taming,dalle,parti,phenaki}. Quantization is a key component to the success of autoregressive models for continuous signals, including images, videos, and audio. The goal of quantization is to provide a compact, discrete representation, which at the same time allows for high-fidelity reconstruction. VQ-VAEs~\citep{vqvae} demonstrated impressive reconstruction quality at low bitrates in various domains and serve as the underlying quantizer for many approaches.

\soundstream{}~\citep{soundstream} is a universal neural audio codec capable of compressing general audio at low bitrates, while maintaining a high reconstruction quality. To achieve this, \soundstream{} uses residual vector quantization (RVQ), allowing scalability to higher bitrate and quality, without a significant computational cost. More specifically, RVQ is a hierarchical quantization scheme composing a series of vector quantizers, where the target signal is reconstructed as the sum of quantizer outputs. Due to the composition of quantizers, RVQ avoids the exponential blowup in the codebook size as the target bitrate increases. Moreover, the fact that each quantizer is fitted to the residual of coarser quantizers introduces a hierarchical structure to the quantizers, where coarser levels are more important for high-fidelity reconstruction. This property is desirable for generation, since the past context can be defined by only attending to the coarse tokens. Recently, \soundstream{} was extended by EnCodec~\citep{defossez2022highfi} to higher bitrates and stereophonic audio. In this work, we rely on \soundstream{} as our audio tokenizer, since it can reconstruct 24 kHz music at 6 kbps with high fidelity. 

\subsection{Generative Models for Audio}

Despite the challenge of generating high-quality audio with long-term consistency, a series of approaches have recently tackled the problem with some success. Jukebox~\citep{jukebox}, for example, proposes a hierarchy of VQ-VAEs at various time resolutions to achieve high temporal coherence, but the generated music displays noticeable artifacts. PerceiverAR~\citep{perceiverAR}, on the other hand, proposes to model a sequence of \soundstream{} tokens autoregressively, achieving high-quality audio, but compromising the long-term temporal coherence. 

Inspired by these approaches, \audiolm{}~\cite{audiolm} addresses the trade-off between coherence and high-quality synthesis by relying on a hierarchical tokenization and generation scheme. Concretely, the approach distinguishes between two token types: (1)~\emph{semantic} tokens that allow the modeling of long-term structure, extracted from models pretrained on audio data with the objective of masked language modeling; (2)~\emph{acoustic} tokens, provided by a neural audio codec, for capturing fine acoustic details.
 This allows \audiolm{} to generate coherent and high-quality speech as well as piano music continuations without relying on transcripts or symbolic music representations.

{\model} builds on top of \audiolm{} with three important additional contributions: (1)~we condition the generation process on a descriptive text, (2)~we show that the conditioning can be extended to other signals such as melody, and (3)~we model a large variety of long music sequences beyond piano music (from \textit{drum'n'bass} over \textit{jazz} to \textit{classical music}).

\subsection{Conditioned Audio Generation}
\looseness=-1
Generating audio from a text description (such as \textit{``whistling with laughter in the background''}) has recently been tackled by several works. DiffSound~\citep{yang2022diffsound} uses CLIP~\citep{clip} as the text encoder and applies a diffusion model to predict the quantized mel spectrogram features of the target audio based on the text embeddings. AudioGen~\citep{audiogen} uses a T5~\citep{t5} encoder for embedding the text, and an autoregressive Transformer decoder for predicting target audio codes produced by EnCodec~\citep{defossez2022highfi}. Both approaches rely on a modest amount of paired training data such as AudioSet~\citep{audioset} and AudioCaps~\citep{audiocaps} (totalling less than 5k~hours after filtering).

 Closer to {\model}, there are also works focusing on music generation conditioned on text. In Mubert~\cite{Mubert}, the text prompt is embedded by a Transformer, music tags which are close to the encoded prompt are selected and used to query the song generation API. Based on the selected tags, Mubert generates a combination of sounds, which in turn were generated by musicians and sound designers. This is in contrast to Riffusion~\cite{riffusion}, which fine-tunes a Stable Diffusion model~\cite{Rombach_2022_CVPR} on mel spectrograms of music pieces from a paired music-text dataset. We use both Mubert and Riffusion as baselines for our work, showing that we improve the audio generation quality and adherence to the text description.

\looseness=-1
Symbolic representations of music (e.g., MIDI) can also be used to drive the generative process as a form of strong conditioning, as demonstrated by~\citet{huang2018music, hawthorne2018wave2midi2wave, engel2020ddsp}. {\model} enables a more natural and intuitive way of providing a conditioning signal, for example through a hummed melody, which can also be combined with a text description.

\subsection{Text-Conditioned Image Generation}

Precursor to text-conditioned audio synthesis are the text-conditioned image generation models, which made significant progress in quality due to architectural improvements and the availability of massive, high-quality paired training data. Prominent Transformer-based autoregressive approaches include \citet{dalle, parti}, while \citet{nichol2021glide, rombach2022high, imagen} present diffusion-based models. The text-to-image approaches have been extended to generating videos from a text prompt~\citep{wu2022nuwa, hong2022cogvideo, phenaki,ho2022video}.

The closest to our approach among these works is {\dalletwo}~\citep{dalle-2}. In particular, similarly to the way {\dalletwo} relies on CLIP~\citep{clip} for text encoding, we also use a joint music-text embedding model for the same purpose. In contrast to  {\dalletwo}, which uses a diffusion model as a decoder, our decoder is based on {\audiolm}. Furthermore, we also omit the prior model mapping text embeddings to music embeddings, such that the {\audiolm}-based decoder can be trained on an audio-only dataset and the music embedding is simply replaced during inference by the text embedding.

\subsection{Joint Embedding Models for Music and Text} \label{subsec:music-text-joint-models}
\mulan~\cite{mulan} is a music-text joint embedding model consisting of two embedding towers, one for each modality. The towers map the two modalities to a shared embedding space of 128~dimensions using contrastive learning, with a setup similar to~\citep{clip,wav2clip}.
The text embedding network is a BERT \cite{bert} pre-trained on a large corpus of text-only data, while we use the ResNet-50 variant of the audio tower.

{\mulan} is trained on pairs of music clips and their corresponding text annotations. Importantly, {\mulan} imposes only weak requirements on its training data quality, learning  cross-modal correspondences even when the music-text pairs are only weakly associated.
The ability to link music to unconstrained natural language descriptions makes it applicable for retrieval or zero-shot music tagging. In this work, we rely on the pretrained and frozen model of \citet{mulan}.

\section{Method}
\label{sec:method}

In this section, we describe {\model} and its components. \cref{sec:audio-representation} describes the models that provide the audio representations. Then, we show in \cref{method-training-inference} how we use these representations for text-conditioned music generation.

\begin{figure}[t]
\begin{center}
\centerline{\includegraphics[trim=5cm 4.1cm 5.5cm 3.4cm, clip, width=0.5\textwidth]{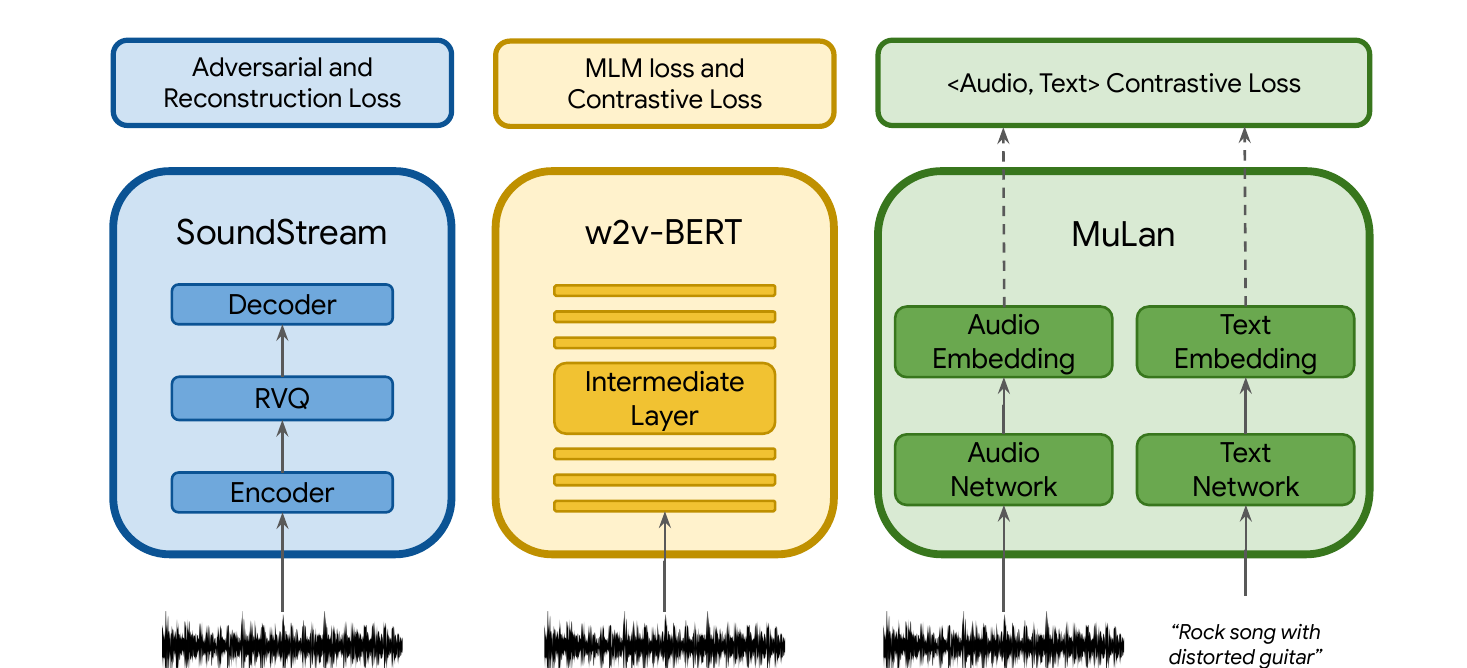}}
\caption{Independent pretraining of the models providing the audio and text representations for {\model}: SoundStream \cite{soundstream}, w2v-BERT \cite{wav2vec-bert}, and MuLan \cite{mulan}.}
\label{fig:models-pretraining}
\end{center}
\end{figure}
\vspace{-4mm}

\subsection{Representation and Tokenization of Audio and Text} \label{sec:audio-representation}
We use three models for extracting audio representations that will serve for conditional autoregressive music generation, which are illustrated in Figure~\ref{fig:models-pretraining}. In particular, by following the approach of {\audiolm}, we use the self-supervised audio representations of {\soundstream}~\citep{soundstream}, as acoustic tokens to enable high-fidelity synthesis, and w2v-BERT~\citep{wav2vec-bert}, as semantic tokens to facilitate long-term coherent generation. For representing the conditioning, we rely on the {\mulan} music embedding during training and the {\mulan} text embedding at inference time. All three of these models are pretrained independently and then frozen, such that they provide the discrete audio and text representations for the sequence-to-sequence modeling.

\paragraph{{\soundstream}.}
\looseness=-1
We use a {\soundstream} model for 24~kHz monophonic audio with a striding factor of 480, resulting in 50~Hz embeddings. The quantization of these embeddings is learned during training by an RVQ with 12 quantizers, each with a vocabulary size of 1024. This results in a bitrate of 6~kbps, where one second of audio is represented by 600~tokens. We refer to these as \emph{acoustic tokens}, denoted by~$\sounds$.

\paragraph{w2v-BERT.}
Similarly to \audiolm, we use an intermediate layer of the masked-language-modeling (MLM) module of a w2v-BERT model with 600M parameters. After pretraining and freezing the model, we extract embeddings from the 7th~layer and quantize them using the centroids of a learned k-means over the embeddings. We use 1024~clusters and a sampling rate of 25~Hz, resulting in 25~\emph{semantic tokens} for every second of audio, denoted by~$\mlm$.

\paragraph{MuLan.}
\looseness=-1
To train {\model}, we extract the representation of the target audio sequence from the audio-embedding network of {\mulan}. Note that this representation is continuous and could be directly used as a conditioning signal in Transformer-based autoregressive models. However, we opt for quantizing the {\mulan} embeddings in such a way that both the audio and the conditioning signal have a homogeneous representation based on discrete tokens,
aiding further research into autoregressively modeling the conditioning signal as well.

\looseness=-1
Since {\mulan} operates on 10-second audio inputs and we need to process longer audio sequences, we calculate the audio embeddings on 10-second windows with 1-second stride and average the resulting embeddings. We then discretize the resulting embedding by applying an RVQ with 12~vector quantizers, each with a vocabulary size of 1024. This process yields 12~{\mulan} audio tokens~$\mulanAudio$ for an audio sequence.
During inference, we use as conditioning the {\mulan} text embedding extracted from the text prompt, and quantize it with the same RVQ as the one used for the audio embeddings, to obtain 12 tokens $\mulanText$.

\looseness=-1
Conditioning on $\mulanAudio$ during training has two main advantages. First, it allows us to easily scale our training data, since we are not limited by the need of text captions. Second, by exploiting a model like {\mulan}, trained using a contrastive loss, we increase the robustness to noisy text descriptions.

\subsection{Hierarchical Modeling of Audio Representations}
\label{method-training-inference}

\begin{figure*}[t!]

    \centering
    \begin{subfigure}[b]{0.59\textwidth}
        \centering
        \includegraphics[trim=4cm 2.8cm 4cm 2.8cm, clip, width=\textwidth]{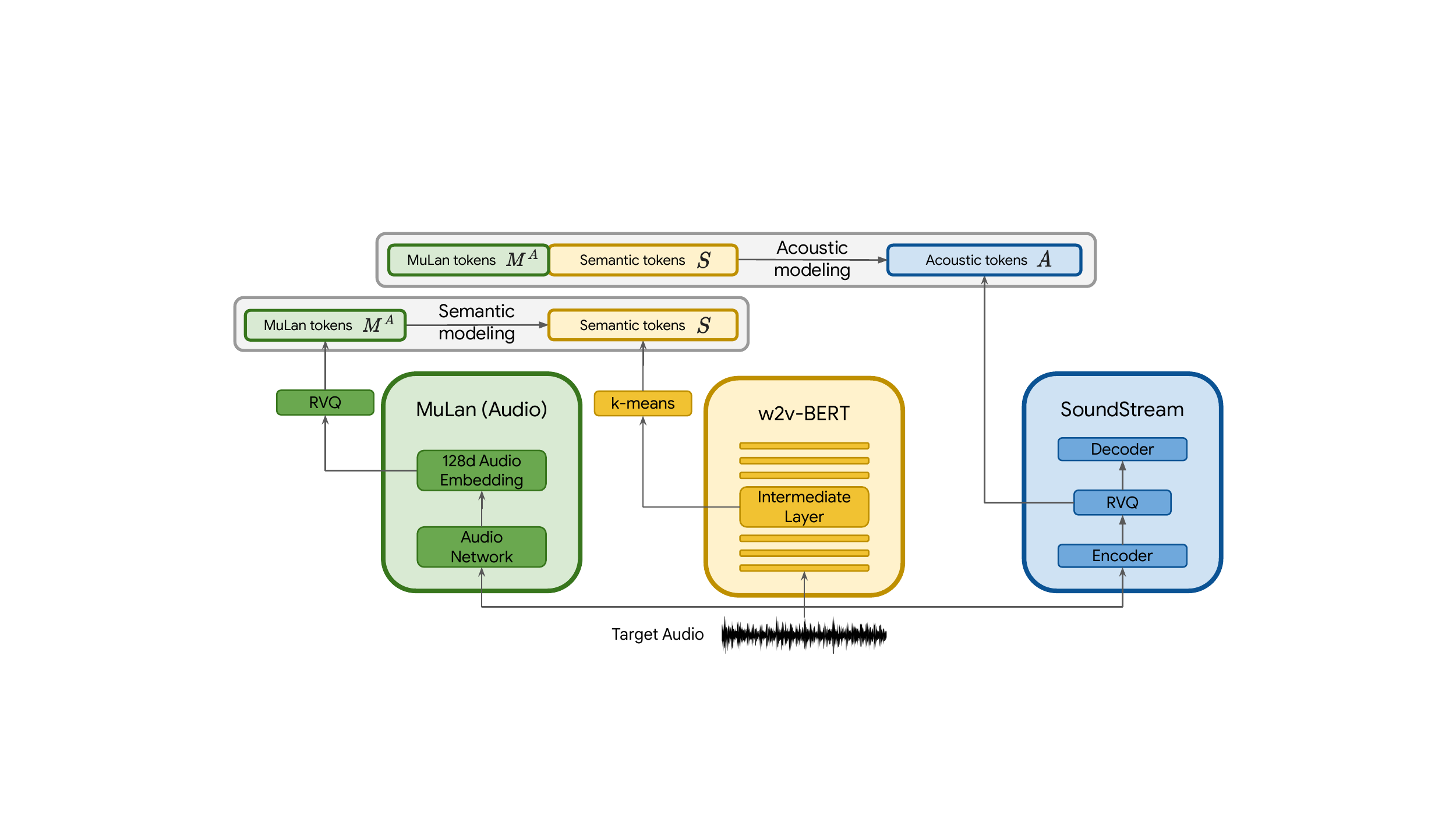}
        \label{fig:training}
    \end{subfigure}
    \hfill
    \begin{subfigure}[b]{0.40\textwidth}
        \centering
        \includegraphics[trim=6.2cm 2.2cm 7.2cm 3.4cm, clip, width=\textwidth]{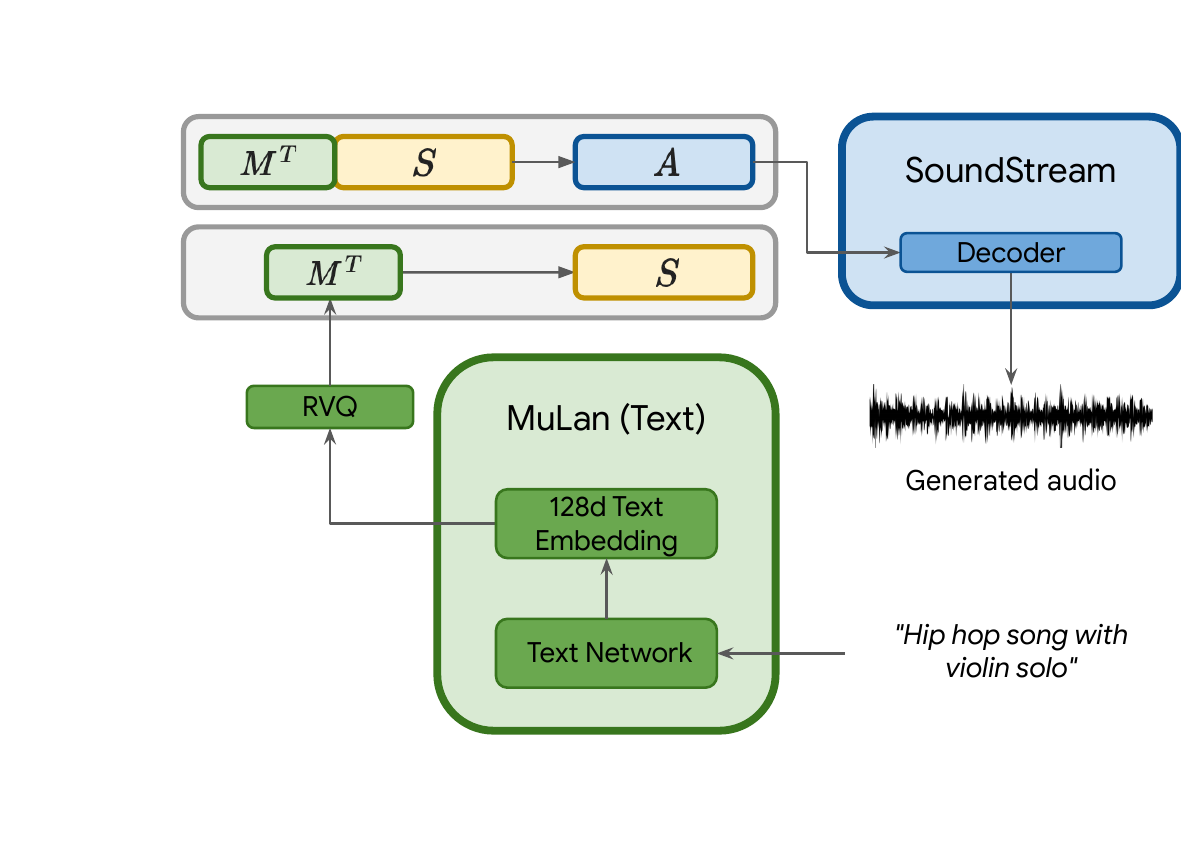}
        \label{fig:inference}
    \end{subfigure}
\vspace{-6mm}
\caption{Left: During training we extract the {\mulan} audio tokens, semantic tokens, and acoustic tokens from the \emph{audio-only} training set. In the semantic modeling stage, we predict semantic tokens using {\mulan} audio tokens as conditioning. In the subsequent acoustic modeling stage, we predict acoustic tokens, given both {\mulan} audio tokens and semantic tokens. Each stage is modeled as a sequence-to-sequence task using decoder-only Transformers.
Right: During inference, we use {\mulan} text tokens computed from the text prompt as conditioning signal and convert the generated audio tokens to waveforms using the SoundStream decoder.}
\label{fig:transformer-pipeline}
\vspace{-1mm}
\end{figure*}

We combine the discrete audio representations presented above with {\audiolm} to achieve text-conditioned music generation. For this, we propose a hierarchical sequence-to-sequence modeling task, where each stage is modeled autoregressively by a separate decoder-only Transformer. The proposed approach is illustrated in \cref{fig:transformer-pipeline}.

The first stage is the \emph{semantic modeling} stage, which learns the mapping from the {\mulan} audio tokens to the semantic tokens $\mlm$, by modeling the distribution $p(\mlm_t\vert\mlm_{<t},\mulanAudio)$, where $t$ is the position in the sequence corresponding to a the time step. 
The second stage is the \emph{acoustic modeling} stage, where the acoustic tokens $\sounds_q$ are predicted conditioned on both the {\mulan} audio tokens and the semantic tokens, modeling the distribution $p(\sounds_t\vert\sounds_{<t},\mlm, \mulanAudio)$.

\looseness=-1
Notably, to avoid long token sequences, {\audiolm} proposed to further split the acoustic modeling stage into a coarse and fine modeling stage. We rely on the same approach, where the coarse stage models the first four levels from the output of the SoundStream RVQ, and the fine stage models the remaining eight --- we refer to~\citet{audiolm} for details.

\section{Experimental Setup}
\label{sec:exp-setup}

\subsection{Models}
We use decoder-only Transformers for modeling the semantic stage and the acoustic stages of {\audiolm}.
The models share the same architecture, composed of 24~layers, 16~attention heads, an embedding dimension of 1024, feed-forward layers of dimensionality~4096, dropout of~0.1, and relative positional embeddings~\cite{t5}, resulting in 430M~parameters per stage. 

\subsection{Training and Inference}

By relying on pretrained and frozen {\mulan}, we need audio-only data for training the other components of {\model}. We train {\soundstream} and w2v-BERT on the Free Music Archive (FMA) dataset~\cite{fma}, whereas the tokenizers and the autoregressive models for the semantic and acoustic modeling stages are trained on a dataset containing five million audio clips, amounting to 280k~hours of music at 24~kHz. Each of the stages is trained with multiple passes over the training data. We use 30~and 10-second random crops of the target audio for the semantic stage and the acoustic stage, respectively. The {\audiolm} fine acoustic modeling stage is trained on 3-second crops.

During inference, we make use of the joint embedding space between audio and text learned by {\mulan}, that is, we substitute $\mulanAudio$ with $\mulanText$. We then follow the stages described above and obtain $\sounds$ given $\mulanText$.
We use temperature sampling for the autoregressive sampling in all stages, with temperature of 1.0 for the semantic modeling stage, 0.95 and 0.4 for the coarse and fine acoustic modeling stages respectively. These temperature values were chosen based on subjective inspection to provide a good trade-off between diversity and temporal consistency of the generated music.

\subsection{Evaluation Dataset}
To evaluate {\model}, we prepare \dataset, a high-quality music caption dataset, which we make publicly available.\footnote{\href{https://www.kaggle.com/datasets/googleai/musiccaps}{kaggle.com/datasets/googleai/musiccaps}}
This dataset includes 5.5k~music clips from AudioSet~\cite{audioset}, each paired with corresponding text descriptions in English, written by ten professional musicians. For each 10-second music clip, {\datasetshort} provides: (1)~a free-text \emph{caption} consisting of four sentences on average, describing the music and  
(2)~a list of music \emph{aspects}, describing genre, mood, tempo, singer voices, instrumentation, dissonances, rhythm, etc. On average, the dataset includes eleven aspects per clip. See Appendix~\ref{sec:appendix-eval-data} for a few caption and aspect list examples.
    
{\datasetshort} complements AudioCaps~\cite{audiocaps}, as they both contain audio clips from AudioSet with corresponding textual descriptions. However, while AudioCaps contains non-music content, {\datasetshort} focuses exclusively on music and includes highly detailed expert-provided annotations. The examples are extracted from both the train and eval split of AudioSet, covering a diverse distribution of genres, as detailed in Appendix~\ref{sec:appendix-eval-data}. {\datasetshort} also provides a \emph{genre-balanced} split of the data with 1k~examples.

\subsection{Metrics}

We compute different metrics to evaluate {\model}, capturing two important aspects of music generation: the audio quality and the adherence to the text description.

\paragraph{Fréchet Audio Distance (FAD).}
The Fréchet Audio Distance~\cite{fad} is a reference-free audio quality metric, which correlates well with human perception.
Models producing samples with a low FAD score are expected to generate plausible audio. However, the generated samples might not necessarily adhere to the text description provided as conditioning.

We report the FAD based on two audio embedding models, both of which are publicly available: (1)~Trill\footnote{\href{https://tfhub.dev/google/nonsemantic-speech-benchmark/trill/3}{tfhub.dev/google/nonsemantic-speech-benchmark/trill/3}} \cite{trill}, which is trained on speech data, and (2)~VGGish\footnote{\href{https://tfhub.dev/google/vggish/1}{tfhub.dev/google/vggish/1}}, \cite{cnn-for-audio-cls} which is trained on the YouTube-8M audio event dataset \cite{yt8m-data}.
Because of the difference in training data, we expect the models to measure different aspects of the audio quality (speech and non-speech, respectively).

\paragraph{KL Divergence (KLD).}
\looseness=-1
There is a many-to-many relationship between text descriptions and music clips compatible with them. It is therefore not possible to directly compare the generated music with the reference at the level of the audio waveform. To assess the adherence to the input text description, we adopt a proxy method similar to the one proposed in~\citet{yang2022diffsound, audiogen}. Specifically, we use a LEAF~\citep{zeghidour_leaf} classifier trained for multi-label classification on AudioSet, to compute class predictions for both the generated and the reference music and measure the KL~divergence between probability distributions of class predictions. When the KL-divergence is low, the generated music is expected to have similar acoustic characteristics as the reference music, according to the classifier.%

\paragraph{MuLan Cycle Consistency (MCC).}
As a joint music-text embedding model, {\mulan} can be used to quantify the similarity between music-text pairs. We compute the {\mulan} embeddings from the text descriptions in {\dataset} as well as the generated music based on them, and define the MCC metric as the average cosine similarity between these embeddings.

\paragraph{Qualitative evaluation.}
\looseness=-1
Ultimately, we rely on subjective tests to evaluate the adherence of generated samples to the text description. We set up an A-vs-B human rating task, in which raters are presented with the text description and two samples of music generated by two different models, or one model and the reference music.
There are five possible answers: strong or weak preference for A or B, and no preference. The raters are instructed not to take the music quality into account when making their decision, because this aspect of the evaluation is already covered by the FAD metric. 

We consider the output of $n$ different models, in addition to the reference music, thus a total of $n+1$~conditions and $n(n+1)/2$~pairs. To aggregate the results of the pairwise tests and rank conditions, we count the number of ``wins'', that is, how often a condition is  strongly or weakly preferred. The samples are selected from the genre-balanced 1k~subset of our evaluation data.

\paragraph{Training data memorization.}
\looseness=-1
Large language models have the capacity to memorize patterns seen in the training data \cite{trm-extract-train-data}. We adapt the methodology used in~\citet{carlini-memorization} to study the extent to which {\model} might memorize music segments. We focus on the first stage, responsible for semantic modeling. We select $N$~examples at random from the training set. For each example, we feed to the model a prompt which includes the MuLan audio tokens $\mulanAudio$ followed by a sequence of the first $\memorizationpromptlen$ semantic tokens $\mlm$, with $\memorizationpromptlen \in \{0,\ldots,250\}$, corresponding to up to 10~seconds. We use greedy decoding to generate a continuation of 125~semantic tokens (5~seconds) and we compare the generated tokens to the target tokens in the dataset. 
We measure exact matches as the fraction of examples for which generated and target tokens are identical over the whole sampled segment. 

\looseness=-1
In addition, we propose a methodology to detect approximate matches, based on the observation that sequences of seemingly different tokens might lead to acoustically similar audio segments. Namely, we compute the histogram of semantic token counts over the corresponding vocabulary $\{0,\ldots,1023\}$ from both the generated and target tokens, and define a matching cost measure between histograms as follows. First, we compute the distance matrix between pairs of semantic tokens, which is populated by the Euclidean distances between the corresponding k-means centroids used to quantize w2v-BERT to semantic tokens (see Section~\ref{sec:audio-representation}). Then, we solve an optimal transport problem to find the matching cost between a pair of histograms using the  Sinkhorn algorithm~\cite{sinkhorn}, considering only the sub-matrix corresponding to non-zero token counts in the two histograms. To calibrate the threshold used to determine whether two sequences might be approximate matches, we construct negative pairs by permuting the examples with target tokens and measure the empirical distribution of matching costs for such negative pairs. We set the match threshold $\matchingthreshold$ to $0.85$, which leads to less than 0.01\% false positive approximate matches. 

\section{Results}
\label{results}

\looseness=-1
We evaluate {\model} by comparing it with two recent baselines for music generation from descriptive text, namely Mubert~\cite{Mubert} and Riffusion~\cite{riffusion}. 
In particular, we generate audio by querying the Mubert API,\footnote{\href{https://github.com/MubertAI}{github.com/MubertAI} (accessed in Dec 2022 and Jan 2023)} and by running inference on the Riffusion model.\footnote{\href{https://github.com/riffusion/riffusion-app}{github.com/riffusion/riffusion-app} (accessed on Dec 27, 2022)}
We perform our evaluations on {\dataset}, the evaluation dataset we publicly release together with this paper.

\paragraph{Comparison to baselines.}
\label{results/comparison}

\cref{table:quant-eval} reports the main quantitative and qualitative results of this paper. In terms of audio quality, as captured by the FAD metrics, on FAD$_{\text{VGG}}$ {\model} achieves better scores than Mubert and Riffusion. On FAD$_{\text{Trill}}$, {\model} scores similarly to Mubert (0.44 vs. 0.45) and better than Riffusion (0.76).
We note that, according to these metrics, {\model} is capable of generating high-quality music comparable to Mubert, which relies on pre-recorded sounds prepared by musicians and sound designers.
In terms of faithfulness to the input text description, as captured by KLD and MCC, {\model} achieves the best scores, suggesting that it is able to capture more information from the text descriptions compared to the baselines.

We further supplement our evaluation of text faithfulness with a human listening test. 
Participants are presented with two 10-second clips and a text caption, and asked which clip is best described by the text of the caption on a 5-point Likert scale. 
We collect 1200 ratings, with each source involved in 600 pair-wise comparisons. Table~\ref{table:quant-eval} reports the total number of ``wins'', that is, counting how often the human raters preferred a model in a side-by-side comparison. 
{\model} is clearly preferred over both baselines, while there is still a measurable gap to the ground truth reference music. Full details of the listening study can be found in Appendix~\ref{sec:qualitative_appendix}.

Listening to examples in which the ground truth was preferred over \model{} reveals the following patterns:
(1)~captions are extremely detailed, referring to more than five instruments or describing non musical aspects such as \textit{``wind, people talking``}; (2)~captions describe temporal ordering of the audio being played; (3)~negations are used, which are not well captured by \mulan.

Overall, we conclude that: (1)~our approach is able to capture fine-grained information from the rich free-text captions of \dataset{}; (2)~the KLD and MCC metrics provide a quantitative measure of the faithfulness to the text description, which is in accordance with the human rating study.

\begin{table}
\setlength{\tabcolsep}{4pt}
\caption{Evaluation of generated samples using captions from the {\dataset} dataset. Models are compared in terms of audio quality, by means of Fréchet Audio Distance (FAD), and faithfulness to the text description, using Kullback–Leibler Divergence (KLD) and  MuLan Cycle Consistency (MCC), and counts of wins in pairwise human listening tests (Wins).}
\label{table:quant-eval}
\vskip 0.15in
\begin{center}
\scriptsize
\begin{sc}
\begin{tabular}{lccccc}
\toprule
Model & FAD$_{\text{Trill}}$ $\downarrow$ & FAD$_{\text{VGG}}$ $\downarrow$ & KLD $\downarrow$ & MCC $\uparrow$ & Wins $\uparrow$ \\
\midrule

Riffusion & 0.76 & 13.4 & 1.19 & 0.34 & 158 \\
Mubert & 0.45 & 9.6 & 1.58 & 0.32 & 97 \\
\model{} & 0.44 & 4.0 & 1.01 & 0.51 & 312 \\
\midrule
\datasetshort{} & - & - & - & - & 472 \\
\bottomrule
\end{tabular}
\end{sc}
\end{center}
\vskip -0.1in
\end{table}

\paragraph{Importance of semantic tokens.}
\label{results/semantic-tokens}
To understand the usefulness of decoupling semantic modeling from acoustic modeling, we train a Transformer model which directly predicts coarse acoustic tokens from MuLan tokens, by modeling $p(\sounds_t\vert{\sounds_{<t}},\mulanAudio)$.
We observe that while the FAD metrics are comparable (0.42~FAD$_{\text{Trill}}$ and 4.0~FAD$_{\text{VGG}}$), KLD and MCC scores worsen when removing the semantic modeling stage. In particular the KLD score increases from 1.01 to 1.05, and the MCC score decreases from 0.51 to 0.49, indicating that semantic tokens facilitate the adherence to the text description.
We also confirm this qualitatively by listening to the samples. In addition, we observe degradation in long term structure.

\paragraph{Information represented by audio tokens.}
\label{results/semantic-vs-acoustic}
We conduct additional experiments to study the information captured by the semantic and the acoustic tokens. In the first study, we fix the {\mulan} text tokens as well as the semantic tokens, running the acoustic modeling stage multiple times to generate several samples. In this case, by listening to the generated music, it is possible to observe that the samples are diverse, yet they tend to share the same genre, rhythmical properties (e.g., drums), and part of the main melody. They differ in terms of specific acoustic properties (e.g., level of reverb, distortion) and, in some cases, different instruments with a similar pitch range can be synthesized in different examples. In the second study, we fix only the {\mulan} text tokens and generate both the semantic and acoustic tokens. In this case, we observe a much higher level of diversity in terms of melodies and rhythmic properties, still coherent with the text description. We provide samples from this study in the accompanying material. 

\begin{figure}[t]
\begin{center}
    \centerline{\includegraphics[clip, width=0.42\textwidth]{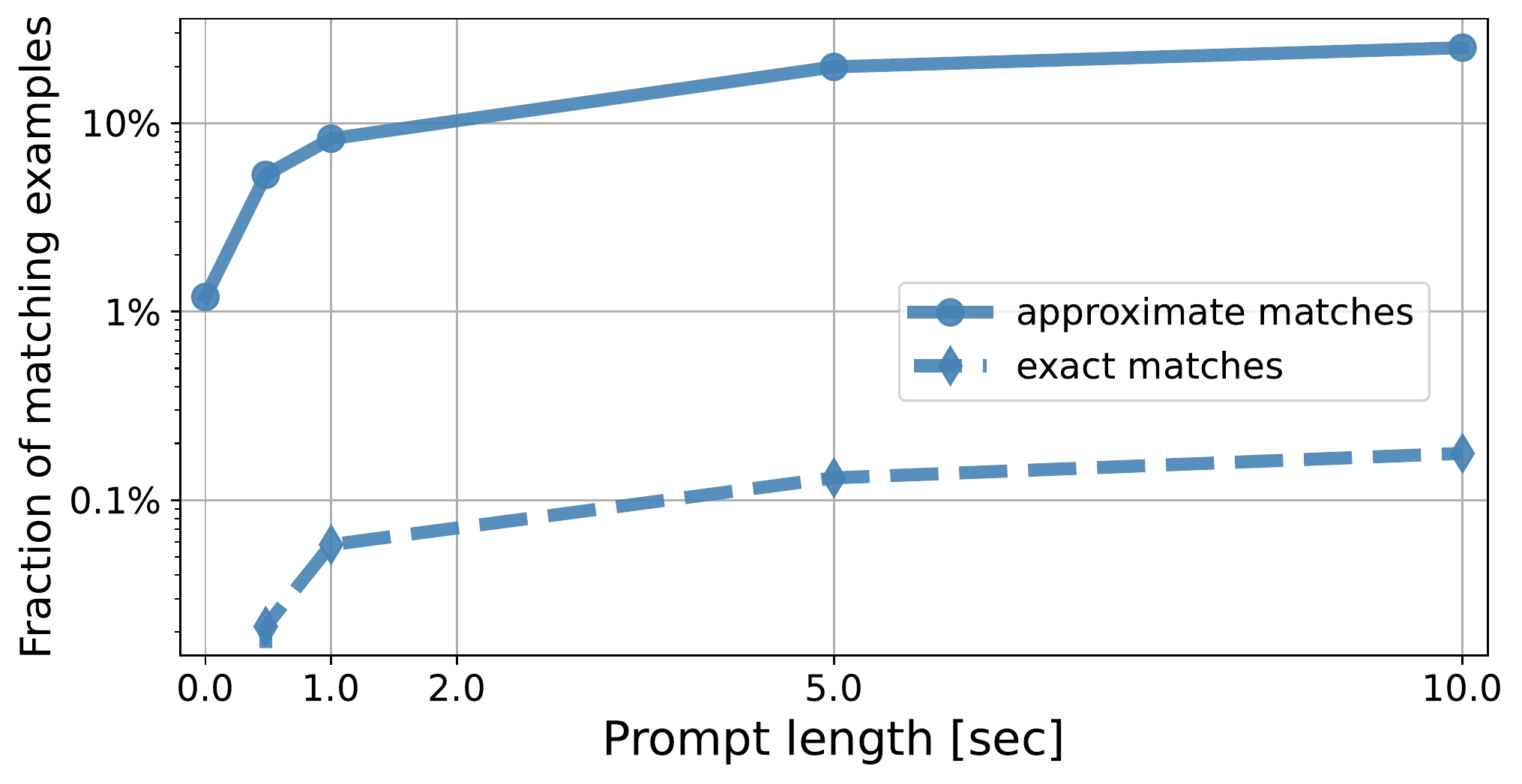}}
    \vspace{-2mm}
    \caption{Memorization results for the semantic modeling stage. We compare the semantic tokens generated for 5~seconds of audio to corresponding tokens in the training set, considering exact and approximate matches.}
    \label{fig:memorization}
    \vspace{-6mm}
\end{center}
\end{figure}

\paragraph{Memorization analysis.}
\label{results/memorization}
\looseness=-1
Figure~\ref{fig:memorization} reports both exact and approximate matches when the length of the semantic token prompt is varied between 0~and 10~seconds. We observe that the fraction of exact matches always remains very small ($< 0.2\%$), even when using a 10 second prompt to generate a continuation of 5 seconds. Figure~\ref{fig:memorization} also includes results for approximate matches, using $\matchingthreshold = 0.85$. We can see a higher number of matches detected with this methodology, also when using only MuLan tokens as input (prompt length $\memorizationpromptlen=0$) and the fraction of matching examples increases as the length of the prompt increases. We inspect these matches more closely and observe that those with the lowest matching score correspond to sequences characterized by a low level of token diversity. Namely, the average empirical entropy of a sample of 125 semantic tokens is 4.6 bits, while it drops to 1.0 bits when considering sequences detected as approximate matches with matching score less than~0.5. We include a sample of approximate matches obtained with $\memorizationpromptlen=0$ in the accompanying material. Note that acoustic modeling carried out by the second stage introduces further diversity in the generated samples, also when the semantic tokens match exactly. 

\section{Extensions}
\paragraph{Melody conditioning.}
\label{melody-conditioning}
We extend {\model} in such a way that it can generate music based on both a text description and a melody, which is provided in the form of humming, singing, whistling, or playing an instrument. 
This requires extending the conditioning signal in a way that captures the target melody. To this end, we create a synthetic dataset composed of audio pairs with matching melodies but different acoustics. To create such pairs, we use different versions of the same music clip, such as covers, instrumentals, or vocals. Additionally, we acquire data pairs of people humming and singing. %
We then train a joint embedding model such that when two audio clips contain the same melody, the corresponding embeddings are close to each other. For implementation details we refer to Appendix~\ref{appendix:melody-conditioning}. 

To extract the melody conditioning for {\model}, we quantize the melody embeddings with RVQ, and concatenate the resulting token sequences with the {\mulan} audio tokens $\mulanAudio$.
During inference, we compute melody tokens from the input audio clip and concatenate them with the {\mulan} text tokens $\mulanText$.
Based on this conditioning, {\model} can successfully generate music which follows the melody contained in the input audio clip, while adhering to the text description.

\paragraph{Long generation and story mode.}
\label{story-mode}
\looseness=-1
In {\model}, generation is autoregressive in the temporal dimension which makes it possible to generate sequences longer than those used during training. In practice, the semantic modeling stage is trained on sequences of 30 seconds. To generate longer sequences, we advance with a stride of 15~seconds, using 15~seconds as prefix to generate an additional 15~seconds, always conditioning on the same text description. With this approach we can generate long audio sequences which are coherent over several minutes.

With a small modification, we can generate long audio sequences while changing the text description over time. Borrowing from ~\citet{phenaki} in the context of video generation, we refer to this approach as \emph{story mode}. Concretely, we compute $\mulanText$ from multiple text descriptions and change the conditioning signal every 15 seconds. The model generates smooth transitions which are tempo consistent and semantically plausible, while changing music context according to the text description.

\section{Conclusions}
\label{sec:conclusions}

We introduce {\model}, a text-conditioned generative model that produces high-quality music at 24~kHz, consistent over several minutes, while being faithful to the text conditioning signal. We demonstrate that our method outperforms baselines on {\dataset}, a hand-curated, high-quality dataset of 5.5k~music-text pairs prepared by musicians.

Some limitations of our method are inherited from {\mulan}, in that our model misunderstands negations and does not adhere to precise temporal ordering described in the text. Moreover, further improvements of our quantitative evaluations are needed. Specifically, since MCC also relies on {\mulan}, the MCC scores are favorable to our method.

Future work may focus on lyrics generation, along with improvement of text conditioning and vocal quality. Another aspect is the modeling of high-level song structure like introduction, verse, and chorus. Modeling the music at a higher sample rate is an additional goal.

\section{Broader Impact}
\label{sec:broader-impact}

{\model} generates high-quality music based on a text description, and thus it further extends the set of tools that assist humans with creative music tasks. However, there are several risks associated with our model and the use-case it tackles. The generated samples will reflect the biases present in the training data, raising the question about appropriateness for music generation for cultures underrepresented in the training data, while at the same time also raising concerns about cultural appropriation. 

\looseness=-1
We acknowledge the risk of potential misappropriation of creative content associated to the use-case. In accordance with responsible model development practices, we conducted a thorough study of memorization, adapting and extending a methodology used in the context of text-based LLMs, focusing on the semantic modeling stage. We found that only a tiny fraction of examples was memorized exactly, while for $1\%$ of the examples we could identify an approximate match. We strongly emphasize the need for more future work in tackling these risks associated to music generation --- we have no plans to release models at this point.

\bibliography{references}
\bibliographystyle{icml2023}

\newpage
\appendix
\onecolumn
\section{{\dataset} Dataset}
\label{sec:appendix-eval-data}

Together with this paper, we release {\dataset}, a high-quality music caption dataset.\footnote{\href{https://www.kaggle.com/datasets/googleai/musiccaps}{kaggle.com/datasets/googleai/musiccaps}}
This dataset includes music clips from AudioSet~\cite{audioset}, paired with corresponding text descriptions in English.
It contains a total of 5,521~examples, out of which 2,858~are from the AudioSet eval and 2,663~from the AudioSet train split.
We further tag 1,000~examples as a balanced subset of our dataset, which is balanced with respect to the genres of the music contained. All examples in the balanced subset are from the AudioSet eval split.

Examples of free text captions:
\begin{itemize}[noitemsep]
    \item \textit{``This folk song features a male voice singing the main melody in an emotional mood. This is accompanied by an accordion playing fills in the background. A violin plays a droning melody. There is no percussion in this song. This song can be played at a Central Asian classical concert.''}
    \item \textit{``This is a live recording of a keyboardist playing a twelve bar blues progression on an electric keyboard. The player adds embellishments between chord changes and the piece sounds groovy, bluesy and soulful.''}
    \item \textit{``A synth is playing an arpeggio pluck with a lot of reverb rising and falling in velocity. Another synth sound is playing pads and a sub bassline. This song is full of synth sounds creating a soothing and adventurous atmosphere. This song may be playing at a festival during two songs for a buildup.''}
    \item \textit{``A low sounding male voice is rapping over a fast paced drums playing a reggaeton beat along with a bass. Something like a guitar is playing the melody along. This recording is of poor audio-quality. In the background a laughter can be noticed. This song may be playing in a bar.''}
    \item \textit{``The electronic music features a section that repeats roughly every two seconds. It consists of a beat that's made of a kick drum and claps. A buzzing synth sets the pulsation of the music by playing once every two beats. The whole music sounds like a loop being played over and over. Towards the end of the excerpt a crescendo-like buzzing sound can be heard, increasing the tension.''}
\end{itemize}

Examples of aspect lists:
\begin{itemize}[noitemsep]
    \item \textit{``pop, tinny wide hi hats, mellow piano melody, high pitched female vocal melody, sustained pulsating synth lead, soft female vocal, punchy kick, sustained synth bass, claps, emotional, sad, passionate''}
    \item \textit{``amateur recording, finger snipping, male mid range voice singing, reverb''}
    \item \textit{``backing track, jazzy, digital drums, piano, e-bass, trumpet, acoustic guitar, digital keyboard song, medium tempo''}
    \item \textit{``rubab instrument, repetitive melody on different octaves, no other instruments, plucked string instrument, no voice, instrumental, fast tempo''}
    \item \textit{``instrumental, white noise, female vocalisation, three unrelated tracks, electric guitar harmony, bass guitar, keyboard harmony, female lead vocalisation, keyboard harmony, slick drumming, boomy bass drops, male voice backup vocalisation''}
\end{itemize}

\newpage

\begin{figure}[H]
\begin{center}
    \centerline{\includegraphics[clip, width=0.7\textwidth]{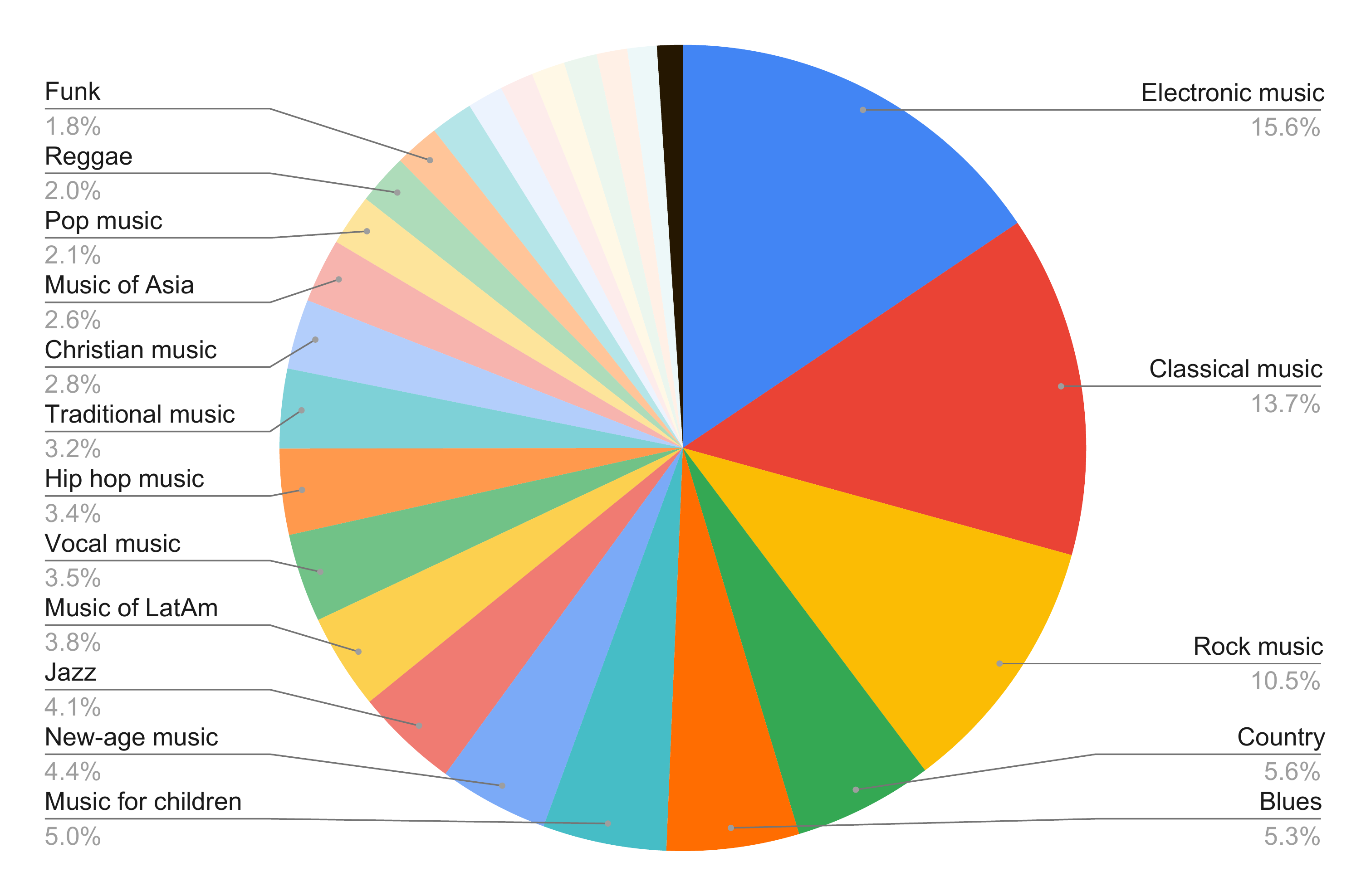}}
    \caption{Genre distribution of all 5.5k~examples of {\datasetshort}, according to an AudioSet classifier.}
    \label{fig:eval-set-genre-dist}
\end{center}
\end{figure}

\begin{figure}[H]
\begin{center}
    \centerline{\includegraphics[clip, width=0.7\textwidth]{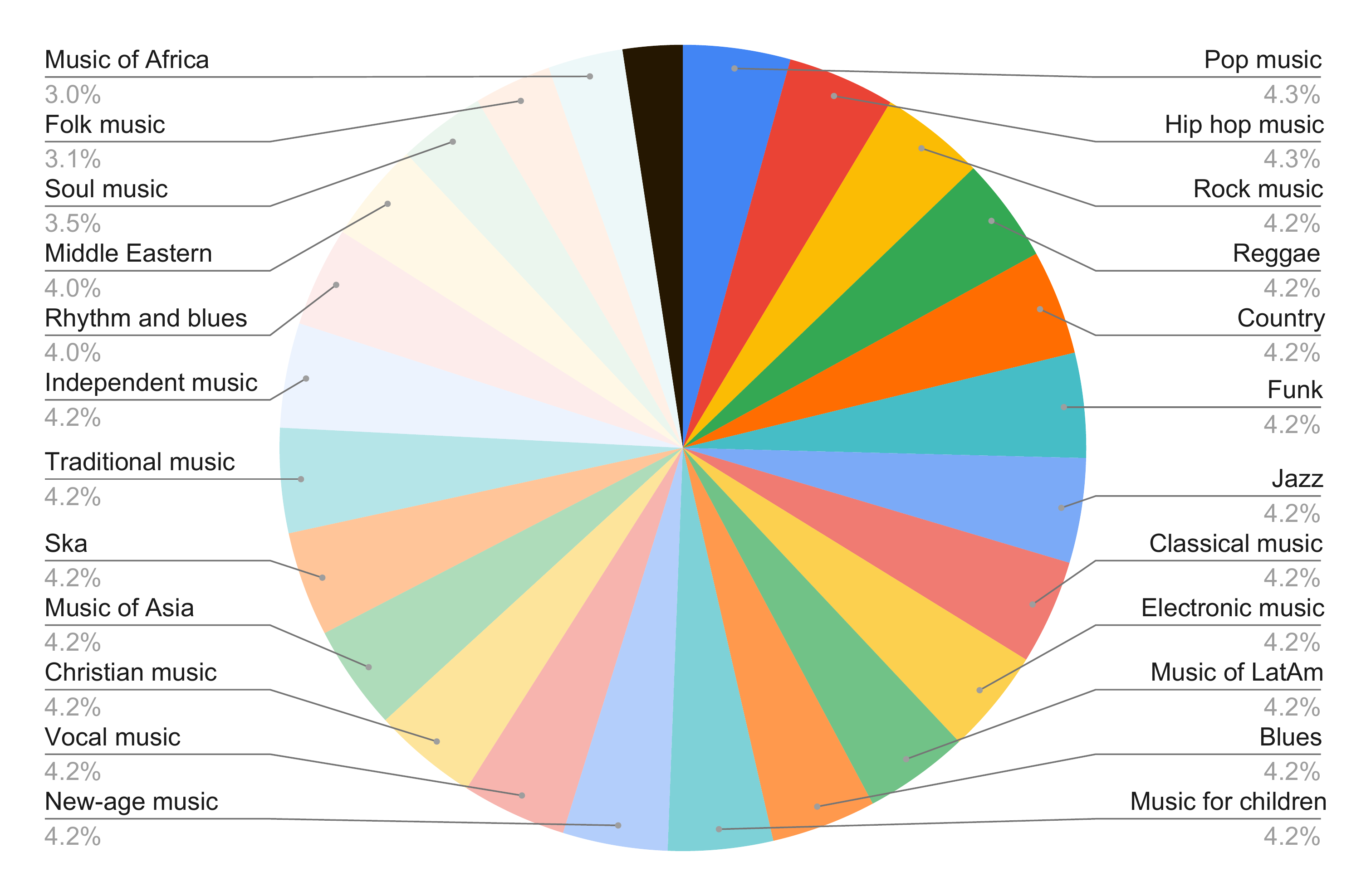}}
    \caption{Genre distribution of a balanced 1k~example subset of {\datasetshort}, according to an AudioSet classifier.}
    \label{fig:eval-set-genre-dist-1k}
\end{center}
\end{figure}

\newpage
\section{Qualitative Evaluation}
\label{sec:qualitative_appendix}

\begin{figure}[H]
\begin{center}
    \centerline{\includegraphics[clip, width=0.7\textwidth]{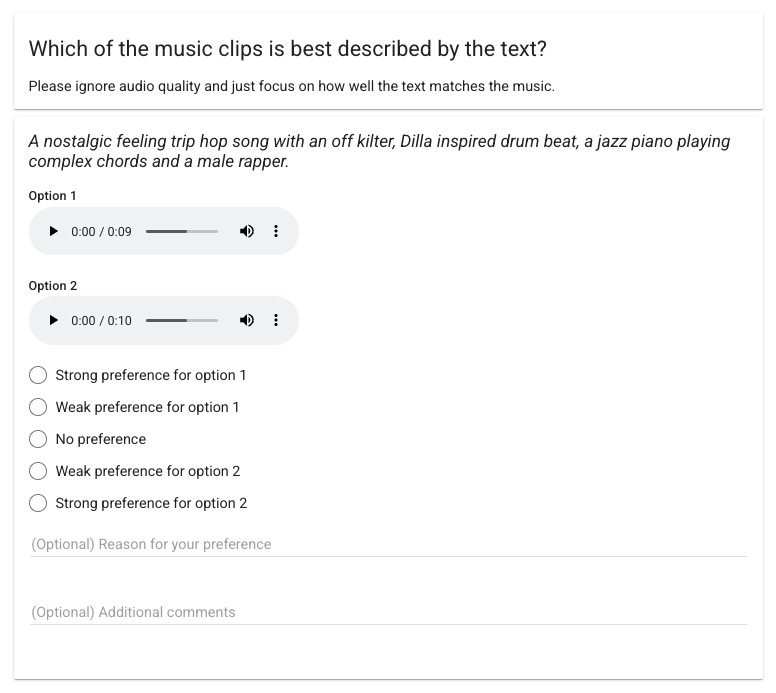}}
    \caption{User interface for the human listener study.}
    \label{fig:listener-ui}
\end{center}
\end{figure}

Participants in the listening test were presented with two 10-second clips and a text caption, and asked which clip is best described the text of the caption on a 5-point Likert scale. 
They were also instructed to ignore audio quality and focus just on how well the text matches the music (similar to \mulan{} score).
Figure~\ref{fig:listener-ui} shows the user interface presented to raters.

\begin{figure}[H]
\begin{center}
    \centerline{\includegraphics[clip, width=1.0\textwidth]{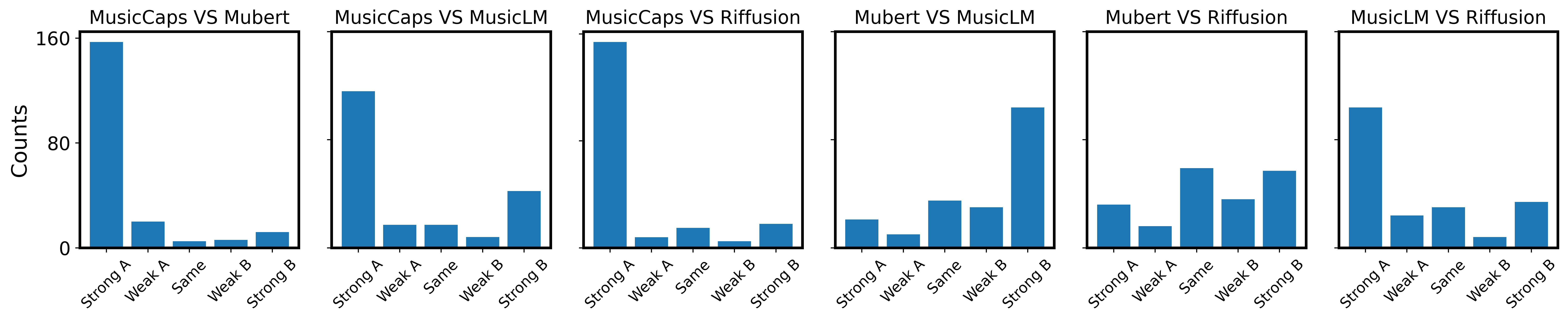}}
    \caption{Pairwise comparisons from the human listener study. Each pair is compared on a 5-point Likert scale. Raters had a decisive model preference in all cases except Mubert vs. Riffusion.}
    \label{fig:pairwise-comparison}
\end{center}
\end{figure}

\begin{figure}[H]
\begin{center}
    \centerline{\includegraphics[clip, width=0.5\textwidth]{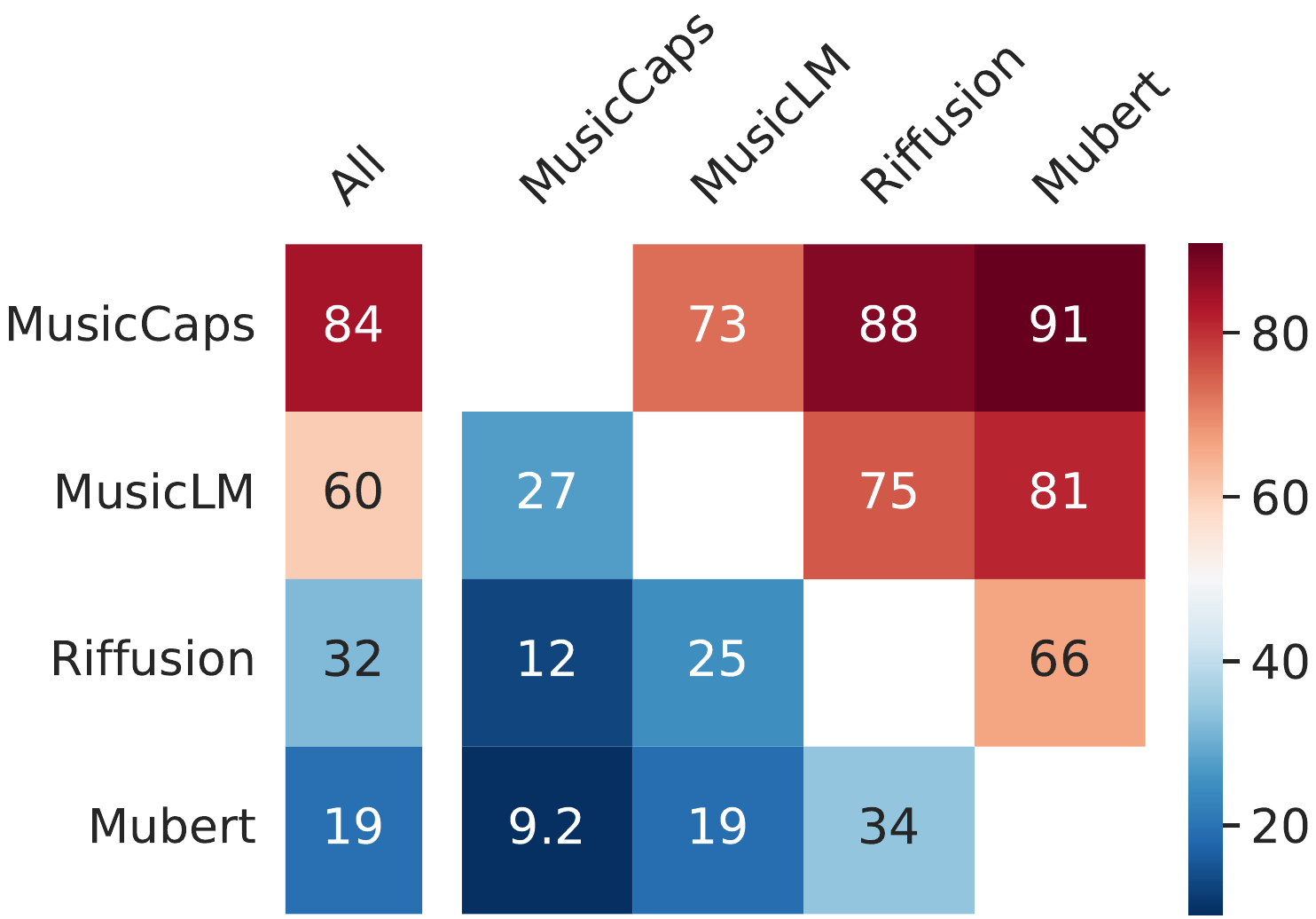}}
    \caption{Win percentage from the human listener study. Each row indicates the \% of times listeners found the music to better match the caption from that system to those from any other system (first column, $N = 1200$) and each system individually (other columns, $N = 600$). The ground truth data (\dataset{}) clearly is the best match to the captions, but followed closely by \model{}, which even beats the ground truth in 27\% of comparisons.}
    \label{fig:listener-confusion}
\end{center}
\end{figure}

We collected 1200 ratings, with each source involved in 600 pair-wise comparisons. 
Figures~\ref{fig:pairwise-comparison} and \ref{fig:listener-confusion} show the granular results of pairwise comparisons between the models.
According to a post-hoc analysis
using the Wilcoxon signed-rank test with Bonferroni correction (with $p < 0.01/15$), the orderings shown in Figure~\ref{fig:listener-confusion} from raters are all statistically significant. 

\section{Melody Conditioning} \label{appendix:melody-conditioning}

We provide here implementation details of the model used for conditioning the music generation on melody.
The model is based on a small ViT~\citep{vit} composed of 12 layers, 6 attention heads, embedding dimension of 512 and feed-forward layer of dimension 1024. The input to the model are the temporal frames of the mel spectrogram of the audio.
We use semi-hard triplet loss \citep{facenet} to train the melody embedding model to generate 192 dimensional embeddings for each 4 seconds of audio.
The model learns to generate embeddings which are representative of a melody while being invariant to acoustic properties related to the instruments being played. This is particularly advantageous, since this representation is complementary to the representation learned by the \mulan{} embeddings. Hence, our melody embeddings and the \mulan{}  can be jointly and complementarily used for conditioning the music generation process.
During training, we consider input audio with a duration of 10 seconds. We extract three melody embeddings, with a hop length of 3 seconds, discretize each of them to tokens with residual vector quantization (RVQ) and concatenate the resulting token sequences with the {\mulan} audio tokens $\mulanAudio$. We use an RVQ composed of 24 quantizers, each with a vocabulary size of 512.

\end{document}